\documentclass[11pt]{article}\def\today{December 15, 2004}
\usepackage{amssymb}
\usepackage{amsmath}
\usepackage{theorem}
\usepackage{latexsym}
\reversemarginpar
\theoremheaderfont{\bf} 
\theorembodyfont{\sl}
\newtheorem{theo}{Theorem}[section]
{\theorembodyfont{\rm} \newtheorem{defi}[theo]{Definition}}
{\theorembodyfont{\rm} \newtheorem{exa}[theo]{Example}}
{\theorembodyfont{\rm} \newtheorem{rem}[theo]{Remark}}
\newtheorem{prop}[theo]{Proposition}

\newtheorem{lemma}[theo]{Lemma}
{\theorembodyfont{\rm}}
{\theorembodyfont{\rm}}
\newenvironment{proof}{{\sc Proof:}}{\mbox{}\hfill$\Box$\par}
\newcommand{\eqnref}[1]{~\mbox{$(${\rm \ref{#1}}$)$}}
\newcommand{\Section}[1]{\section{#1}\setcounter{equation}{0}}

\newcommand{\junk}[1]{}
\newcommand{\DS}{\displaystyle}

\newcommand{\N}{{\mathbb N}}
\newcommand{\F}{{\mathbb F}}

\newcommand{\cC}{{\mathcal C}}

\newcommand{\cS}{{\mathcal S}}

\newcommand{\CC}{convolutional code}
\newcommand{\CCC}{cyclic convolutional code}
\newcommand{\ve}[1]{\mbox{$\varepsilon^{(#1)}$}}

\newcommand{\rank}{\mbox{\rm rank}\,}
\newcommand{\ord}{\mbox{\rm ord}}

\newcommand{\AutF}{\mbox{${\rm Aut}_{\mathbb F}$}}
\newcommand{\im}{\mbox{\rm im}\,}
\renewcommand{\mod}{\mbox{\rm mod}\,}

\newcommand{\nkd}{\mbox{$(n,k,\delta)$}}
\newcommand{\dist}{\mbox{\rm dist}}
\newcommand{\wt}{\mbox{\rm wt}}

\newcommand{\T}{\mbox{$\!^{\sf T}$}}

\newcommand{\ideal}[1]{\mbox{$\langle{#1}\rangle$}}
\newcommand{\Flaurent}{\mbox{$\F(\!(z)\!)$}}
\newcommand{\Azs}{\mbox{$A[z;\sigma]$}}

\newcommand{\p}{\mbox{$\mathfrak{p}$}}

\newcounter{abc}
\newcounter{def}
\newenvironment{romanlist}{\begin{list}{(\roman{abc})\hfill}{\usecounter{abc}
     \topsep-1.4ex \labelwidth.7cm \leftmargin.7cm \labelsep0cm
     \rightmargin0cm \parsep0ex \itemsep.6ex
     \partopsep1.6ex}}{\end{list}}
\newenvironment{alphalist}{\begin{list}{(\alph{abc})\hfill}{\usecounter{abc}
     \topsep-1.4ex \labelwidth.7cm \leftmargin.7cm \labelsep0cm
     \rightmargin0cm \parsep0ex \itemsep.6ex
     \partopsep01.6ex}}{\end{list}}
\newenvironment{arabiclist}{\begin{list}{(\arabic{abc})\hfill}{\usecounter{abc}
     \topsep-1.4ex \labelwidth.7cm \leftmargin.7cm \labelsep0cm
     \rightmargin0cm \parsep0ex \itemsep.6ex
     \partopsep1.6ex}}{\end{list}}

\title{A Class of One-Dimensional MDS Convolutional Codes}
\date\today
\author{Heide Gluesing-Luerssen\footnote{
       University of Groningen, Department of Mathematics, P.~O.~Box 800,
       9700 AV Groningen, The Netherlands; gluesing@math.rug.nl}
       \ and Barbara Langfeld\footnote{
       Kombinatorische Geometrie (M9),
       Zentrum Mathematik, Technische Universit\"at M\"unchen, Boltzmannstr.~3,
       85747 Garching bei M\"unchen, Germany;
       langfeld@ma.tum.de}
       }
\begin{document}
\maketitle

\begin{abstract}
\noindent 
A class of one-dimensional convolutional codes will be presented. 
They are all MDS codes, i.~e., have the largest distance among all one-dimensional 
codes of the same length~$n$ and overall constraint length~$\delta$.
Furthermore, their extended row distances are computed, and they increase with 
slope~$n-\delta$.
In certain cases of the algebraic parameters, we will also derive parity check matrices of 
Vandermonde type for these codes.
Finally, cyclicity in the convolutional sense of~\cite{GS04} will 
be discussed for our class of codes.
It will turn out that they are cyclic if and only if the field element used in the 
generator matrix has order~$n$.
This can be regarded as a generalization of the block code case.

\end{abstract}



\Section{Introduction}

The main task of coding theory is the construction of powerful codes.
This applies equally well to block codes and convolutional codes.
In either case codes are required to have good error-correcting properties, i.~e.\ a 
large distance, and an efficient decoding algorithm. 

In block coding theory this goal has been achieved best by the class of Reed-Solomon codes 
along with their efficient algebraic decoding algorithm.
These codes are in particular MDS (maximum distance separable), meaning that they have the
largest distance possible among all codes with the same length and dimension.
On the other hand there are convolutional codes, and despite their frequent and successful 
use in engineering practice, their mathematical theory is still in the beginnings. 
The algebraic theory of this class of codes has been initiated with the 
paper~\cite{Fo70} of Forney and has seen a considerable development ever since.

In particular, quite some efforts have been made in the area of constructing convolutional 
codes with large distance. 
The first group of the according papers appeared in the seventies of the last century.
In \cite{Ju73,MCJ73,Ju75} quasi-cyclic block codes have been used in order to construct
convolutional codes with good distance. 
The relation between the weights of the block codewords and the convolutional codewords 
is made by the weight-retaining property.
This topic has been resumed later on in~\cite{SGR01} where the ideas have been used 
to construct MDS convolutional codes with (almost) arbitrary algebraic parameters. 
Other more recent attempts of constructing good convolutional codes try to impose additional
algebraic structure on the convolutional codes themselves.  
In~\cite{PPS04} methods from algebraic geometry are used in order to construct 
convolutional codes of Goppa type.
In the paper~\cite{HRS03} system theoretic methods are used in order to analyze 
codes with optimal column distances.
Finally, in~\cite{GS04} ideas from the seventies~\cite{Pi76,Ro79} have been resumed in 
order to impose a type of cyclicity on convolutional codes. 
The investigations of these cyclic convolutional codes have been continued 
in~\cite{GS03,GL03}. 
We will explain the notion of cyclicity later in Section~4 of this paper. 
At this moment we restrict ourselves to mentioning that cyclicity for convolutional 
codes is a more general notion than just the natural invariance of the code under 
cyclic shift.

In the present paper we will combine the two main lines mentioned above. 
We will present a class of one-dimensional codes that are not only MDS but also 
have extended row distances increasing with slope $n-\delta$ (where~$n$ is the length of the 
code and~$\delta$ the overall constraint length). 
We will also compare the required field size needed for the construction with the field sizes
of other constructions known in the literature. 
It will turn out that our field sizes are smaller for many parameters than what 
has been used before. 
For one set of parameters the field size is even only one above the theoretic minimum. 
In addition to these distance computations and field size investigations, we will also 
discuss the algebraic structure of these codes.
As it turns out, for certain algebraic parameters the presented codes are cyclic in the 
sense mentioned above. 
In this case the codes can in fact be regarded as a generalization of (one-dimensional) 
Reed-Solomon codes. 
They even have a polynomial parity check matrix of Vandermonde 
type, showing that this class of codes are closely related to some of the codes given 
in~\cite{PPS04}.

The paper is organized as follows.
In the rest of the introduction we will collect the preliminaries about convolutional codes.
In Section~2 we will present the class of codes via their generator matrices along with 
their (extended row) distances and compare the field size to results from the literature. 
In Secton~3 various parity check matrices with Vandermonde structure are presented. 
In Section~4 we will introduce the notion of cyclicity for convolutional codes as it 
has been investigated in~\cite{GS04}. 
We will show that in a certain (to be expected) case our codes are cyclic, and we will 
present various representations of the codes. 
We will close with some open problems.

We end this introduction with the basic notions of convolutional coding theory.
Convolutional codes are certain submodules of~$\F[z]^n$, where~$\F$ is a finite
field.
Before presenting the definition we wish to recall that each submodule~$\cS$ of
$\F[z]^n$ is free and therefore can be written as
\[
   \cS=\im G:=\big\{uG\,\big|\, u\in\F[z]^k\big\}
\]
where~$k$ is the rank of~$\cS$ and $G\in\F[z]^{k\times n}$ is a matrix
containing a basis of~$\cS$. 
Hence, the matrix~$G$ is unique up to left multiplication by a matrix from
$Gl_k(\F[z])$. 
Moreover, by resorting to the Smith normal form one easily shows that 
$G$ is right-invertible, i.~e., $G\tilde{G}=I_k$ for some matrix 
$\tilde{G}\in\F[z]^{n\times k}$, if and only if the submodule $\im G$ is a direct 
summand of the module $\F[z]^n$. This in turn is equivalent to the existence of a matrix
$H\in\F[z]^{(n-k)\times n}$ such that $\im G=\ker H\T:=\{v\in\F[z]^n\mid vH\T=0\}$.
Using the theory of polynomial matrices it is easily seen that we may assume~$H$ to be
right-invertible. 
Then it is unique up to left multiplication by a matrix from $Gl_{n-k}(\F[z])$.
Obviously, the matrix~$H$ generates the {\em dual module}, i.~e.,  
$\im H=\cS^{\perp}:=\{w\in\F[z]^n\mid w v\T=0\text{ for all }v\in\cS\}$. 

This makes all of the following notions well-defined.

\begin{defi}\label{D-CC}
Let $\F$ be any finite field.
A {\em convolutional code\/} $\cC\subseteq\F[z]^n$ with (algebraic) 
parameters $(n,k,\delta)$ is a submodule of the form $\cC=\im G$, where 
$G\in\F[z]^{k\times n}$ is a right-invertible matrix such that 
$\delta=\max\{\deg\gamma\mid \gamma\text{ is a }k\text{-minor of }G\}$.
We call~$G$ a {\em generator matrix\/} of the code. 
The number~$n$ is called the {\em length}, ~$k$ is the {\em dimension}, and~$\delta$ is 
called the {\em overall constraint length\/} of the code. 
Each right-invertible matrix $H\in\F[z]^{(n-k)\times n}$ satisfying $\cC=\ker H\T$ is 
called a {\em parity check matrix\/} of~$\cC$. 
\end{defi}
Thus, the convolutional codes of length~$n$ are the direct summands of~$\F[z]^n$.
It is worth mentioning that a code with overall constraint length zero can be 
regarded as a block code. 
In the coding literature a right invertible matrix is often called
{\em basic\/}~\cite[p.~730]{Fo70} or
{\em delay-free and non-catastrophic}, see~\cite[p.1102]{McE98}.
Sometimes in the literature convolutional codes are defined as subspaces of
the vector space $\Flaurent^n$ of vector valued Laurent series over~$\F$, see for
instance~\cite{McE98} and~\cite{Fo70}.
However, as long as one restricts to right invertible generator matrices
it does not make a difference whether one works in the context of infinite message 
and codeword sequences or finite ones, see also~\cite{RSY96,Ro01}.

The most important concept for a code is its distance.
It measures the error-correcting capability, hence the quality, of the code.
The definition of the distance of a convolutional code is straightforward.
For a polynomial vector $v=\sum_{j=0}^N v_j z^j\in\F[z]^n$
the {\em weight\/} is defined as $\wt(v)=\sum_{j=0}^N\wt(v_j)$, where $\wt(v_j)$ 
denotes the usual Hamming weight of $v_j\in\F^n$.
Then the {\em (free) distance\/} of a code $\cC\subseteq\F[z]^n$ with generator matrix
$G\in\F[z]^{k\times n}$ is given as
\[
   \dist(\cC):=\min\{\wt(v)\mid v\in\cC,\;v\not=0\}=
   \min\big\{\wt(uG)\,\big|\, u\in\F[z]^k,\;u\not=0\big\}.
\]
Just like for block codes there exist quite some bounds on the distance of 
convolutional codes.
One of them is the {\em generalized Singleton bound\/}~\cite[Thm.~2.2]{RoSm99}.
It states that the distance~$d$ of a code with parameters \nkd\ over any field satisfies
\begin{equation}\label{e-MDS}
   d\leq S\nkd:=(n-k)\Big(\Big\lfloor\frac{\delta}{k}\Big\rfloor+1\Big)+\delta+1.
\end{equation}
Notice that $S(n,k,0)=n-k+1$ which is the well-known Singleton bound for block codes.
Like for block codes we call a code~$\cC$ with $\dist(\cC)=S\nkd$ an MDS code 
(maximum distance separable), see~\cite[Def.~2.5]{RoSm99}.
Observe also that\eqnref{e-MDS} can easily be seen if $k=1$.
Indeed, $S(n,1,\delta)=n(\delta+1)$ is clear since in this case each generator matrix, 
being a codeword itself, obviously has weight at most $n(\delta+1)$.

\Section{A class of one-dimensional MDS codes}

In this section we present a construction of one-dimensional MDS convolutional codes.
The distance will be computed straightforwardly. 
We will then compare our results with constructions known from the literature.
Thereafter we will also compute the extended row distances.
We will derive that they are increasing with a slope of~$n-\delta$.

\begin{theo}\label{T-MDS}
Let $n\in\N$ and $0\leq\delta\leq n-1$ and
let $q$ be a prime power such that $n\leq q-1$. Put $\F:=\F_q$ and
choose an element $\alpha\in\F$ such that $\ord(\alpha)\geq n$.
Define
\begin{equation}\label{e-Gmatrix}
   G:=\sum_{\nu=0}^{\delta}z^\nu
   \begin{pmatrix}1&\alpha^\nu&\alpha^{2\nu}&\ldots&\alpha^{(n-1)\nu}\end{pmatrix}
   \in\F[z]^{1\times n}
\end{equation}
and let $\cC:=\im G\subseteq\F[z]^n$.
Then $G$ is right invertible, i.~e., the submodule~$\cC$ is a convolutional code, and
$\dist(\cC)=n(\delta+1)$.
In other words, $\cC$ is an MDS code with parameters $(n,1,\delta)$.
\end{theo}
Notice that for $\delta=0$ the code is simply the $n$-fold repetition (block) code 
over~$\F$ and the assertions are obvious.

\begin{proof}
In order to show that $G$ is right invertible, we have to prove
that the entries of the matrix~$G$ are coprime.
In other words, it needs to be proven that the polynomials
\[
  \sum_{\nu=0}^{\delta}z^\nu,\ \sum_{\nu=0}^{\delta}(\alpha z)^\nu,\
  \sum_{\nu=0}^{\delta}(\alpha^2 z)^\nu,
  \ldots,\
  \sum_{\nu=0}^{\delta}(\alpha^{n-1}z)^\nu
\]
have no common root in any extension field~$\hat{\F}$ of~$\F$. In order to
see this, assume $\beta\in\hat{\F}$ is such a common root. Then
$\beta, \alpha\beta,\ldots,\alpha^{n-1}\beta$ are roots of
$\sum_{\nu=0}^{\delta}z^\nu$. Since $\beta\not=0$ and
$\ord(\alpha)\geq n$, these numbers are pairwise different
and $\delta<n$ leads to a contradiction.
\\[.7ex]
Next we will prove that $\dist(\cC)=n(\delta+1)$.
To this end put
$G_\nu:=(1,\,\alpha^\nu,\,\alpha^{2\nu},\,\ldots,\alpha^{(n-1)\nu})$
for $\nu=0,\ldots,\delta$.
Let $u=\sum_{i=0}^t u_iz^i\in\F[z]$, where $t\geq0$ and
$u_0\not=0\not=u_t$, and put $uG=:v=\sum_{i=0}^{\delta+t}v_iz^i$.
Defining $G_\nu:=0$ for $\nu<0$ and $\nu>\delta$, we have
\[
    v_\nu=(u_0,\ldots,u_t)\tilde{G}_\nu, \text{ where }
    \tilde{G}_\nu=\begin{pmatrix}
               G_\nu\\ G_{\nu-1}\\ \vdots\\G_{\nu-t}\end{pmatrix}
\]
for $\nu=0,\ldots,\delta+t$.
Notice that for $\nu\leq\delta$ the first row, $G_\nu$, of $\tilde{G}_\nu$ is nonzero
while for $\nu\geq t$ the last row, $G_{\nu-t}$, is nonzero.
Since $u_0\not=0\not=u_t$ this will provide us with a good estimate of the
weight of $v_\nu$ for these indices.
In order to see this, note that for each index~$\nu$ the nonzero rows of
$\tilde{G}_\nu$ are consecutive and form a matrix of the type
\[
  R:=\begin{pmatrix} 1&\alpha^{s+r}&\alpha^{2(s+r)}&\cdots&\alpha^{(n-1)(s+r)}\\
                     \vdots&\vdots&\vdots& &\vdots\\
                     1&\alpha^{s+1}&\alpha^{2(s+1)}&\cdots&\alpha^{(n-1)(s+1)}\\
                     1&\alpha^s&\alpha^{2s}&\cdots&\alpha^{(n-1)s}
    \end{pmatrix}
\]
where $0\leq s\leq s+r\leq\delta$.
Since $\ord(\alpha)\geq n>\delta$, the block code $\im R\subseteq\F^n$ is MDS,
that is,
\begin{equation}\label{e-MDSblock}
   \dist(\im R)=n-r.
\end{equation}
This will now be used for counting the weight of the vectors~$v_\nu$ for
$\nu\in\{0,\ldots,\delta,t,\ldots,\delta+t\}$.
\\
\underline{1.~case:} $t>\delta$
\\
In this case the indices $0,\ldots,\delta,t,\ldots,\delta+t$ are all different
and we have
\begin{equation}\label{e-tildeG}
  \tilde{G}_\nu=\begin{pmatrix}G_\nu\\G_{\nu-1}\\\vdots\\G_0\\0\\\vdots\\0\end{pmatrix}
  \text{ for }\nu=0,\ldots,\delta
  \text{ and }
  \tilde{G}_\mu=\begin{pmatrix}0\\\vdots\\0\\G_{\delta}\\G_{\delta-1}\\\vdots\\G_{\mu-t}
  \end{pmatrix}\text{ for }\mu=t,\ldots,\delta+t
\end{equation}
and all displayed rows~$G_{\ell}$ are nonzero.
Thus, using\eqnref{e-MDSblock},
\begin{equation}\label{e-estim1a}
  \wt(v_\nu)\geq n-\nu\text{ for }\nu=0,\ldots,\delta
  \text{ and }
  \wt(v_\mu)\geq n-(\delta+t-\mu)\text{ for }\mu=t,\ldots,\delta+t,
\end{equation}
and therefore
\begin{equation}\label{e-estim1b}
  \wt(v)\geq 2\big(n+(n-1)+\ldots+(n-\delta)\big)
  =2n(\delta+1)-\delta(\delta+1)\geq n(\delta+1)
\end{equation}
where the last inequality follows from $\delta<n$.
\\
\underline{2.~case:} $t\leq\delta$
\\
In this case we consider the indices
$0,\ldots,\delta,\delta+1,\ldots,\delta+t$.
For $\nu=0,\ldots,t$ and for $\mu=\delta+1,\ldots,\delta+t$ the matrices
$\tilde{G}_\nu$ and $\tilde{G}_\mu$ are as in\eqnref{e-tildeG}, while for
$\lambda=t+1,\ldots,\delta$ we have
\[
 \tilde{G}_\lambda=\begin{pmatrix}
    G_\lambda\\G_{\lambda-1}\\\vdots\\G_{\lambda-t}\end{pmatrix}
\]
and, again, all block rows~$G_{\ell}$ of~$\tilde{G}_\lambda$ are nonzero.
From this and\eqnref{e-MDSblock} we obtain
\begin{align}
  \wt(v)&\geq\big(n\!+\!(n\!-\!1)\!+\!\ldots\!+\!(n\!-\!t)\big)\!+\!(\delta\!-\!t)(n\!-\!t)
             \!+\!\big((n\!-\!t\!+\!1)\!+\!(n\!-\!t\!+\!2)\!+\!\ldots\!+\!n\big)\nonumber\\
        &=2\Big(tn-\sum_{i=0}^{t-1}i\Big)+(\delta-t+1)(n-t)
         =n(\delta+1)+t(n-\delta)
          \geq n(\delta+1).\label{e-estim2}
\end{align}
This concludes the proof.
\end{proof}

The proof above also shows that $uG$ with $u\in\F^k\backslash\{0\}$, i.~e., the nonzero
constant multiples of~$G$, are the only codewords having weight $n(\delta+1)$.
Indeed, the inequality in\eqnref{e-estim1b} is always strict and the last inequality
in\eqnref{e-estim2} is strict for all $t>0$.

\begin{rem}\label{R-largedelta}
It is not hard to see that the matrix~$G$ in\eqnref{e-Gmatrix} is also right-invertible for 
all $\delta\geq n$ for which $\ord(\alpha)\nmid\delta+1$. 
Examples show that these codes often have a large distance, too, but are not MDS in 
general.
However, we can not provide any general result in this case. 
\end{rem}

We would like to comment on the field size required for the construction of the
MDS codes in Theorem~\ref{T-MDS}. 
In \cite[Lemma~1]{Ju75} and \cite[Thm.~3.7]{GS03} it has been shown that if $\cC$ is an 
$(n,1,\delta)$-MDS code over $\F_q$ then $q\geq\delta+1$.
In Theorem~\ref{T-MDS} the field size~$q$ satisfies $q\geq n+1\geq\delta+2$.
Thus, in the case $n=\ord(\alpha)=q-1$ and $\delta=n-1$ our 
field size is just one above the lower bound given above.
As to our knowledge it is not known in general whether there exist
$(n,1,n-1)$-MDS codes over $\F_n$ (in the case where~$n$ is a prime power).
\\
We also would like to compare our results with previous constructions of MDS codes.
In~\cite{Ju75} MDS codes with parameters $(n,1,\delta)$ for certain combinations have been 
constructed. However, these combinations are different from ours. 
For instance, the result in \cite[Thm.~p.~580]{Ju75} does not contain the case of 
$(q-1,1,q-2)$-MDS codes over~$\F_q$ and no $(q-1,1,q-3)$-MDS codes over $\F_q$ where $q>5$.
On the other hand, the construction of that theorem allows the construction of a
$(17,1,20)$-MDS code over~$\F_{32}$ which is not part of our Theorem~\ref{T-MDS}.
In~\cite{SmRo98} a construction of $(n,1,\delta)$-MDS codes is given over fields
$\F_q$ where $q>\delta n+1$.
Except for the case $\delta=1$ this is a considerably bigger field size than ours
where $q\geq n+1$.
However, the construction in \cite{SmRo98} works for all~$\delta$ and not just for 
$\delta<n$.
Another construction of MDS codes is given in~\cite{SGR01}. Therein, MDS codes with 
(almost) arbitrary parameters $(n,k,\delta)$ are constructed over fields~$\F_q$ of size
$q\geq\frac{\delta n^2}{k(n-k)}+2$. 
The construction is based on cyclic block codes with large distance.
In the case $k=1$ this again amounts to a considerably bigger field than in our construction.
Alternatively, one can also see directly that our codes are not derived from good 
cyclic block codes in the sense of~\cite{SGR01}, i.~e., the polynomial 
$g=\sum_{j=1}^n z^{j-1}G_j(z^n)$ derived from $G=\big(G_1(z),\ldots,G_n(z)\big)$ does not 
generate a good cyclic block code in general.

We want to go into more details about the weight distribution of these codes and therefore
give also lower bounds for the extended row distances. 
The extended row distances have been introduced in
\cite[p.~541]{JPB90} and are very closely related to the trellis structure of the code
and thus to its performance. 
Details on the importance of these distance parameters can be found 
in~\cite{JPB90}\footnote{The row distances, as defined
in~\cite[p.~114]{JoZi99} do not give any further information. They are all equal to the 
free distance $n(\delta+1)$.}.
The $j$th extended row distance amounts to the minimum weight of all paths through the 
state diagram starting at the zero state and which reach the zero state after exactly~$j$ 
steps for the first time.
In other words, it is the minimum weight of all atomic codewords of degree~$j-1$ 
(i.~e., length~$j$) in the sense of~\cite{McE98a}.
The details are also explained in~\cite[Sec.~3.10]{JoZi99}.
In our case where the dimension of the code is~$k=1$, the atomic codewords are easily 
described.
We will confine ourselves to the following property. It follows readily from the fact that 
the last~$\delta$ coefficients of the message $u\in\F[z]$ make up the current state 
in the state diagram.

\begin{lemma}\label{L-atomic}
Let $G\in\F[z]^{1\times n}$ be a right-invertible generator matrix of the code 
$\cC:=\im G\subseteq\F[z]^n$ and let~$G$ have overall constraint length $\delta>0$.
Let $u\in\F[z]$. Then the following are equivalent.
\begin{romanlist}
\item The codeword $uG$ is atomic (i.~e., the associated path through the state diagram 
      does not pass through the zero state except for its starting and end point).
\item The polynomial $u\in\F[z]$ does not have $\delta$ consecutive zero coefficients.
\end{romanlist}
\end{lemma}

Having this property in mind, the $j$th extended row distance of the code $\cC=\im G$ is 
defined to be
\[
  \hat{d}^r_j:=\min\Big\{\wt(uG)\,\Big|\,
        \begin{array}{l} u\in\F[z],\,u_0\not=0,\,\deg u=j-\delta-1,\\
             \text{no~$\delta$ consecutive coefficients of~$u$ are zero}\end{array}\Big\}
  \text{ for all }j\geq\delta+1.
\]
Notice that $\deg(u)=j-\delta-1$ implies $\deg(uG)=j-1$ and thus the associated path has 
length~$j$.
As for the index notation we diverge somewhat from the paper~\cite{JPB90}
where the index~$j$ equals the degree of the associated codewords while in 
our case it reflects the length.

\begin{prop}\label{P-extrowdist}
Let $\cC=\im G\subseteq\F[z]^n$ be the code described in Theorem~\ref{T-MDS}.
Then the extended row distances satisfy
\[
   \hat{d}^r_j\geq(n-\delta)j+\delta(\delta+1)\text{ for all }j\geq\delta+1.
\]
Hence the extended row distances are bounded from below by a linear function with slope 
$n-\delta$.
\end{prop}

Before we prove this result we wish to mention that in a certain sense this result is 
the best one can expect.
As Equation\eqnref{e-nextcoeff} below shows, the ``middle'' coefficients of a codeword 
are contained in an $(n,\delta+1)$-block code. 
The distance of this code is therefore a lower bound for the slope.
In our case this code has optimum distance $n-\delta$, therefore the weight of codewords 
increases at least linearly in the length with slope~$n-\delta$.
However, in specific cases certain constellations of consecutive coefficients of the 
generator matrix might even allow a better row distance. 
After the proof we will present examples for both cases, where the estimate in 
Proposition~\ref{P-extrowdist} is actually an identity and where it is a 
strict inequality.

\noindent{\sc Proof:}
Let $u\in\F[z]$ and $\deg u=j-\delta-1\geq0$.
Then $uG=:v=\sum_{i=0}^{j-1}v_iz^i$ has degree $j-1$ and length~$j$.
\\
If $j-\delta-1\leq\delta$, then\eqnref{e-estim2} shows
$\wt(v)\geq n(\delta+1)+(j-\delta-1)(n-\delta)=\delta(\delta+1)+j(n-\delta)$.
\\
Let now $j-\delta-1>\delta$. From\eqnref{e-estim1a} we have
$\wt\Big(\sum_{i=0}^{\delta}v_iz^i+\sum_{i=j-\delta-1}^{j-1}v_iz^i\Big)
  \geq (2n-\delta)(\delta+1)$.
Thus it remains to consider the coefficients $v_i$ where $i=\delta+1,\ldots,j-\delta-2$.
Since
\begin{equation}\label{e-nextcoeff}
  v_i=\sum_{l=0}^{\delta}u_{i-l}G_l
  =(u_i,u_{i-1},\ldots,u_{i-\delta})\begin{pmatrix}G_0\\G_1\\ \vdots\\G_{\delta}
                                    \end{pmatrix}
\end{equation}
and~$v$ is atomic, the vector $(u_i,u_{i-1},\ldots,u_{i-\delta})$ is nonzero by 
Lemma~\ref{L-atomic}.
Thus $\wt(v_i)\geq n-\delta$ by\eqnref{e-MDSblock} and we obtain
$$
  \wt(v)\geq (2n-\delta)(\delta+1)+(j-2\delta-2)(n-\delta)=(n-\delta)j+\delta(\delta+1).
  \eqno\Box
$$

In the following examples we consider various cases of the  parameters~$n$ and~$\delta$
in Theorem~\ref{T-MDS}. 
We computed the exact weight distribution 
(see \cite[Sec.~3]{McE98a}) of the codes using Maple. 

\begin{exa}\
\begin{arabiclist}
\item Let $n=3,\,\delta=1$ and $\F=\F_4=\{0,1,\alpha,\alpha^2\}$ where $\alpha^2=\alpha+1$. 
      Consider~$G$ as in Theorem~\ref{T-MDS}, that is
      \[
        G=\begin{pmatrix}1+z&1+\alpha z&1+\alpha^2z\end{pmatrix}.
      \]
      In this case one can show that the weight distribution is given by
      (see, e.~g., \cite[Sec.~3.10]{JoZi99} and~\cite{McE98a}) 
      \[
        A(L,W) = 3L^2W^6/(1-3LW^2)=\sum_{j=2}^{\infty}3^{j-1}W^{2+2j}L^j,
      \]
      meaning that all atomic codewords of length~$j$ have weight~$2+2j$ and that there 
      exist~$3^{j-1}$ of such codewords for each~$j\geq2$.
      As a consequence, the estimate in Proposition~\ref{P-extrowdist} is an equality, 
      i.~e., $\hat{d}^r_j=2+2j$.
      One can also present explicitly an atomic codeword of length~$j$ and weight $2+2j$.
      Indeed, it can be shown directly that $\wt\big((\sum_{i=0}^{j-2}z^i)G\big)=2+2j$ 
      for each $j\geq2$.
\item Let $\delta=1,\,\text{char}(\F)=2$ and $n$ be arbitrary. Then it is easy to see that
      $\wt\big((\sum_{i=0}^{j-2}z^i)G\big)=2+j(n-1)$ for each $j\geq2$, 
      hence the estimate in Proposition~\ref{P-extrowdist} is an identity.
\item Let $\delta=2$ and, again, $n=3,\,\F=\F_4$. Then~$G$ as defined in Theorem~\ref{T-MDS} 
      is given by
      \[
          G = \begin{pmatrix}1+z+z^2&1+\alpha z+\alpha^2 z^2&1+\alpha^2z+\alpha z^2
	       \end{pmatrix}.
      \]
      In this case the weight distribution is
      \begin{align*}
        A(L,W)&= 3W^9L^3(1+2LW-2LW^3)/(6L^3W^8\!-\!6L^3W^6\!-\!3L^2W^5\!-\!2LW^3\!-\!LW+1)\\
              &=3W^9L^3+9W^{10}L^4+(9W^{11}+18W^{13}+9W^{14})L^5+O(L^6),
      \end{align*}
      meaning, for instance, that there are $36$ atomic codewords of length five, $9$ of 
      which have weight $11$ and~$14$, respectively, and~$18$ have weight~$13$. 
      Using induction it is easy to see that in the series expansion for each $j\geq3$ the 
      coefficient of~$L^j$ is divisible by $W^{6+j}$ but not by~$W^{7+j}$. 
      Hence, $\hat{d}^r_j=6+j$, and, like in~(2), the estimate in 
      Proposition~\ref{P-extrowdist} is an equality. Again, in this case one has
      $\wt\big((\sum_{i=0}^{j-3}z^i)G\big)=6+j$ for each $j\geq3$.
\item In general however, the inequality for the $j$th extended row distance is not an 
      equality and the growth rate can even be better.
      This happens for instance for $n=3,\,\delta=2$ and $\F=\F_{8}$ with, of course,
      $\ord(\alpha)=7$.
      In this case the weight distribution of the code in Theorem~\ref{T-MDS} is
      \begin{align*}
      A(L,W)=&\; 7\,W^9L^3+(21W^{10}\!+\!28W^{12})L^4+(14W^{12}\!+\!
               126W^{13}\!+\!147W^{14}\!+\!105W^{15})L^5\\
            &+(91W^{14}+\ldots)L^6+(63W^{15}+\ldots)L^7+(28W^{16}+\ldots)L^8\\
	    &+(28W^{17}+\ldots)L^9
	     +(154W^{19}+\ldots)L^{10}+(56W^{20}+\ldots)L^{11}\\
	    &+(56W^{21}+\ldots)L^{12}+(392W^{23}+\ldots)L^{13}+(168W^{24}+\ldots)L^{14}+O(L^{15})
      \end{align*}
      where each sum ``$+\ldots$'' is meant to contain only higher powers of~$W$. 
      This shows that the weight distribution is even better than the lower bound given in
      Proposition~\ref{P-extrowdist}. 
      At least for small~$j$ we have $\hat{d}^r_j>j+\delta(\delta+1)=j+6$.
\end{arabiclist}
\end{exa}

\Section{Parity check matrices with Vandermonde structure}

In this section we will derive two types of parity check matrices for the codes of 
Theorem~\ref{T-MDS} in the case where $\ord(\alpha)=n=\delta+1$, one of them being 
minimal in the sense of \cite[p.~459]{Fo75}.
Both reveal a type of Vandermonde structure for these codes.

\begin{theo}\label{T-Vanderm}
Let $\ord(\alpha)=n$ and consider the matrix 
\[
   H:=\begin{pmatrix}
       z-\alpha^n&z-\alpha^{n-1}&\cdots &z-\alpha^2&z-\alpha\\
       (z-\alpha^n)^2&(z-\alpha^{n-1})^2&\cdots &(z-\alpha^2)^2&(z-\alpha)^2\\
        \vdots&       \vdots           &       &\vdots        &\vdots\\
       (z-\alpha^n)^{n-1}&(z-\alpha^{n-1})^{n-1}&\cdots &(z-\alpha^2)^{n-1}&(z-\alpha)^{n-1}\\
      \end{pmatrix}\in\F[z]^{(n-1)\times n}.
\]
Then
\begin{arabiclist}
\item $H$ is right-invertible,
\item $G H\T=0$ where 
      $G=\sum_{\nu=0}^{n-1}z^\nu
        \begin{pmatrix}1&\alpha^\nu&\alpha^{2\nu}&\ldots&\alpha^{(n-1)\nu}\end{pmatrix}$.
\end{arabiclist}
Hence, in the case where $\ord(\alpha)=n=\delta+1$, the code given in Theorem~\ref{T-MDS} 
has parity check matrix~$H$. 
\end{theo}

The condition $\ord(\alpha)=n$ is necessary for the theorem to be true. 
As can easily be checked the product $GH\T$ is, in general, not zero if $\ord(\alpha)>n$.

\begin{proof}
(1) For $j=1,\ldots,n$ let $H^{(j)}\in\F[z]^{(n-1)\times(n-1)}$ be the submatrix of~$H$ 
obtained by omitting the $j$th column. 
Then, due to the Vandermonde structure of~$H$, we obtain
\[
  \det H^{(j)}=\!\!\!\!\!\prod_{{\nu=1\atop \nu\not=n-j+1}}^n\!\!\!\!(z-\alpha^{\nu})
               \!\!\!\prod_{\nu=1\atop\nu\not=n-j+1}^n
	       \prod_{\mu=\nu+1\atop\mu\not=n-j+1}^n\!\!(z-\alpha^{\mu}-z+\alpha^{\nu})
	      =\!\!\!\!\prod_{{\nu=1\atop \nu\not=n-j+1}}^n\!\!\!\!(z-\alpha^{\nu})
	       \!\!\!\prod_{1\leq\nu<\mu\leq n\atop \nu,\mu\not=n-j+1}\!\!
	                   (\alpha^{\nu}-\alpha^{\mu}).			
\]
Since $\ord(\alpha)=n$, the last factor is nonzero for each~$j$. 
But then the first factors show the coprimeness of
the maximal minors of~$H$, and thus~$H$ is right-invertible~\cite[Thm.~A.1]{McE98}.
\\
(2) Let~$G$ be given as above. 
Then for $j=1,\ldots,n$ the $j$th entry~$G_{j}$ is of the form
\[
  G_j=\sum_{\nu=0}^{n-1}\big(\alpha^{j-1}z\big)^{\nu}
      =\frac{(\alpha^{j-1}z)^n-1}{\alpha^{j-1}z-1}
     =\frac{z^n-1}{\alpha^{j-1}z-1}=\alpha^{n-j+1}\frac{z^n-1}{z-\alpha^{n-j+1}},
\]
where for the last equality we used $\ord(\alpha)=n$. 
Thus,
\begin{equation}\label{e-G}
   G=\Big(\alpha^n\frac{z^n-1}{z-\alpha^n},\,\alpha^{n-1}\frac{z^n-1}{z-\alpha^{n-1}},\ldots,
          \alpha\frac{z^n-1}{z-\alpha}\big).
\end{equation}
Now we can prove $GH\T=0$.
For easier indexing we will write down the sums of the matrix product backwards.
Then we have to show that 
\[
    \sum_{\nu=1}^n\alpha^{\nu}\frac{z^n-1}{z-\alpha^{\nu}}\cdot(z-\alpha^{\nu})^j=0 
    \text{ for }j=1,\ldots,n-1.
\]
This is equivalent to 
\begin{equation}\label{e-GHt}
  \sum_{\nu=1}^n\alpha^{\nu}(z-\alpha^{\nu})^j=0\text{ for }j=0,\ldots,n-2.
\end{equation}
In order to see this, compute
\[
   \sum_{\nu=1}^n\alpha^{\nu}(z-\alpha^{\nu})^{j}
   =\sum_{\nu=1}^n\alpha^{\nu}\sum_{\mu=0}^{j}{j\choose\mu}
                                     z^{j-\mu}(-1)^{\mu}\alpha^{\nu\mu}
   =\sum_{\mu=0}^{j}z^{j-\mu}{j\choose\mu}(-1)^{\mu}\sum_{\nu=1}^n\alpha^{\nu(\mu+1)}.
\]
Notice that for fixed $\mu=0,\ldots,j$ we have $\mu+1\in\{1,\ldots,j+1\}$ and 
$j\leq n-2$ yields $\alpha^{\mu+1}\not=1$ due to $\ord(\alpha)=n$.
Therefore, 
\begin{equation}\label{e-alphpowers}
   \sum_{\nu=1}^n\alpha^{\nu(\mu+1)}=\sum_{\nu=0}^{n-1}\alpha^{\nu(\mu+1)}
   =\frac{\alpha^{(\mu+1)n}-1}{\alpha^{\mu+1}-1}=0 \text{ for all }\mu=0,\ldots,j. 
\end{equation}
This proves the Equations\eqnref{e-GHt} and thus $GH\T=0$.
\end{proof}
One should notice that the parity check matrix~$H$ is highly non-minimal, i.~e., it is 
not a minimal basis for the dual code~$\cC^{\perp}$
(for the notion of minimal basis see \cite[p.~459]{Fo75} or \cite[Sec.~2.5]{JoZi99}).
Obviously the leading coefficient matrix is the all-$1$-matrix and thus has rank~$1$ only. 
This implies non-minimality of~$H$ by \cite[Main Thm.]{Fo75}.
A minimal parity check matrix will be presented at the end of this section.

The reader will have noticed that we did not make use of the Vandermonde parity check 
matrix~$H$ when computing the distances of the codes in the last section.
As to our knowledge no theoretical result is known yet about the distances of convolutional 
codes with Vandermonde generator or parity check matrices.
As an indication that such a relation is not obvious, we would like to mention that
Vandermonde parity check matrices of the form
\[
   H:=\begin{pmatrix}
       (z-\alpha^n)^r&(z-\alpha^{n-1})^r&\cdots &(z-\alpha^2)^r&(z-\alpha)^r\\
       (z-\alpha^n)^{r+1}&(z-\alpha^{n-1})^{r+1}&\cdots &(z-\alpha^2)^{r+1}&(z-\alpha)^{r+1}\\
        \vdots&       \vdots           &       &\vdots        &\vdots\\
       (z-\alpha^n)^s&(z-\alpha^{n-1})^s&\cdots &(z-\alpha^2)^s&(z-\alpha)^s\\
      \end{pmatrix}\in\F[z]^{(s-r+1)\times n}
\]
with $0\leq r\leq s\leq n-1$, i.~e., with fewer rows than 
the matrix in~Theorem~\ref{T-Vanderm}, do not in general lead to good codes, even if 
$\ord(\alpha)=n$.
This can easily be seen by running a few examples using, for instance, Maple.
However, one should also notice the close relation of these matrices to those appearing in
\cite[Exa.~4.1]{PPS04}.
In that paper methods from algebraic geometry are used to construct convolutional codes of
Goppa type. A few examples of such matrices, but with different linear factors in the 
entries, are presented in~\cite{PPS04} which are generator matrices of MDS convolutional 
codes.
A deeper understanding as to whether there is a relation between the distance of the codes 
$\ker H\T$ or $\im H$ and the Vandermonde structure of~$H$ must be considered as one of 
the main tasks in algebraic convolutional coding theory.
It might also have some impact on the possibility of algebraic decoding of these codes.

\begin{rem}\label{R-Vandermonde}
With completely different methods it is possible to prove that also in the general case
$0\leq\delta<n=\ord(\alpha)$, the codes from Theorem~\ref{T-MDS} have a parity check
matrix of a (somewhat modified) Vandermonde type. 
Indeed, in that case such a matrix is given by 
\[
  H:=\begin{pmatrix}
   1&\alpha&\cdots&\alpha^{n-1}\\
   1&\alpha^2&\cdots&\alpha^{2(n-1)}\\
   \vdots&\vdots& & \vdots\\
   1&\alpha^{n-\delta-1}&\cdots&\alpha^{(n-\delta-1)(n-1)}\\
   z-\alpha^n&z-\alpha^{n-1}&\cdots&z-\alpha\\
   (z-\alpha^n)^2&(z-\alpha^{n-1})^2&\cdots&(z-\alpha)^2\\
   \vdots&\vdots& &\vdots\\
   (z-\alpha^n)^{\delta}&(z-\alpha^{n-1})^{\delta}&\cdots&(z-\alpha)^{\delta}
  \end{pmatrix}\in\F[z]^{(n-1)\times n}.
\]
Hence~$H$ is right invertible and satisfies $G H\T=0$.
The proof of this statement needs more detailed methods from the theory of cyclic 
convolutional codes as derived in~\cite{GS04} and will be omitted.
\end{rem}

At the end of this section we will return to the case where $\delta=n-1$ and present a 
minimal parity check matrix, i.~e., a right-invertible matrix with minimal row degrees 
in the sense of \cite[p.~459]{Fo75} or \cite[Sec.~2.5]{JoZi99}.
It shows that the dual code of $\im G$ has Forney index~$1$ 
(counted ($n-1$) times)\footnote{The Forney indices of a code are defined to be the 
row degrees of a minimal generator matrix, see \cite[p.~1081]{McE98}.}
and thus is a compact code in the sense of \cite[Cor.~4.3]{McE98}.

\begin{theo}\label{T-MinParityCheck}
Let again $\ord(\alpha)=n$ and define
\[  
     H_{\it{min}}:=\Big((\alpha^{n-\nu+1})^{j-1}z-(\alpha^{n-\nu+1})^j\Big)_{
               \!\!j=1,\ldots,n-1\atop\nu=1,\ldots,n\quad }
     \in\F[z]^{(n-1)\times n}.
\]
Then $H_{\it{min}}$ is minimal and right-invertible and $G\,H_{\it{min}}^{\sf T}=0$, where~$G$ 
is again as in Theorem~\ref{T-Vanderm}(2).
Hence in the case where $\ord(\alpha)=n=\delta+1$ the matrix $H_{\it{min}}$ is a parity 
check matrix of the code given in Theorem~\ref{T-MDS}.
\end{theo}

\begin{proof}
We use again the representation\eqnref{e-G} for the matrix~$G$.
Writing down the sums of the product $G H_{\it{min}}^{\sf T}$ backwards  again we 
obtain
\[
  \sum_{\nu=1}^n\alpha^{\nu}\frac{z^n-1}{z-\alpha^{\nu}}
   \big((\alpha^{\nu})^{j-1}z-(\alpha^{\nu})^j\big)
  =(z^n-1)\sum_{\nu=0}^{n-1}(\alpha^j)^{\nu}.
\]
But the last expression is zero for all $j=1,\ldots,n-1$ as we have shown 
in\eqnref{e-alphpowers}. 
From this we obtain that $\im H_{\it{min}}\subseteq\ker G\T=(\im G)^{\perp}$.
Hence $H_{\it{min}}=B\hat{H}$, where~$\hat{H}$ is a parity check matrix of the code
$\im G$ and~$B$ is some polynomial matrix. 
By \cite[Thm.~3]{Fo75} the overall constraint length of $\im\hat{H}$ is $n-1$, too.
On the other hand it is seen directly, that, firstly, the matrix $H_{\it{min}}$ has 
full row rank~$k$ and that, secondly, the overall constraint length of 
$\im H_{\it{min}}$ is the sum of the row degrees of $H_{\it{min}}$ because the 
highest coefficient matrix of $H_{\it{min}}$ is a Vandermonde matrix with full row rank,
see also \cite[p.~495]{Fo75}. 
Hence $H_{\it{min}}$ is minimal and both matrices $H_{\it{min}}$ and~$\hat{H}$ have 
overall constraint length~$n-1$.
This shows that $\det(B)\in\F\backslash\{0\}$ and thus $H_{\it{min}}$ is 
right-invertible, too. 
\end{proof}

Notice that $H_{\it{min}}=H_1z-H_0$ where both $H_1$ and $H_0$ have 
Vandermonde structure. 
It is worth mentioning that the dual codes, i.~e., the codes generated by~$H$ or 
$H_{\it{min}}$, are in general not optimal, that is, they are not MDS (this can be
checked by a few examples).
They do not even attain in general the Griesmer bound, see 
\cite[Thm.~3.22]{JoZi99} or \cite[Thm.~3.4]{GS03} for the non-binary case. 
We also wish to point out the slight similarity of our construction with that 
in \cite[pp.~445]{Pi88}. Therein, an MDS code with parity check matrix
of the form $H_1z+H_0$, where $H_0,\,H_1$ are Vandermonde matrices, is presented. 
However, in that construction the code has large dimension $k>\frac{n}{2}$ while in our 
case $k=1$.

\Section{Cyclicity}

In this section we will show that for positive overall constraint length the codes 
given in Theorem~\ref{T-MDS} are cyclic if and only if $\ord(\alpha)=n$.
Cyclic convolutional codes have been studied in detail in~\cite{GS04}.
The first investigations in this direction have been made in the seventies
by Piret~\cite{Pi76} and Roos~\cite{Ro79}.
In both papers it has been shown (with different methods and in different
contexts) that cyclicity of \CC{}s must not be understood in
the usual sense, i.~e. invariance under the cyclic shift, if one wants to
go beyond the theory of cyclic block codes.
As a consequence, Piret suggested a more complex notion of cyclicity which then has been
further generalized by Roos.
In both papers some nontrivial examples of \CCC{}s in
this new sense are presented along with their distances.
All this indicates that the new notion of cyclicity seems to be the
appropriate one in the convolutional case.
Recently, in the paper~\cite{GS04} an algebraic theory of \CCC{}s has been established 
which goes well beyond the results of the seventies.
On the one hand it leads to a nice, yet nontrivial, generalization of the theory of
cyclic block codes, on the other hand it gives a very powerful toolbox for
constructing such codes.
We will now give a very brief introduction into cyclicity for convolutional codes before 
investigating this additional structure for the codes of Theorem~\ref{T-MDS}.

Just like for cyclic block codes we assume from now on that the length~$n$
and the field size~$|\F|$ are coprime. 
Recall that a block code $\cC\subseteq\F^n$ is called cyclic if it is
invariant under the cyclic shift, i.~e.
\begin{equation}\label{e-cs}
  (v_0,\ldots,v_{n-1})\in\cC\Longrightarrow
  (v_{n-1},v_0,\ldots,v_{n-2})\in\cC
\end{equation}
for all $(v_0,\ldots,v_{n-1})\in\F^n$.
This is the case if and only if~$\cC$ is an ideal in
the quotient ring
\begin{equation}\label{e-A}
     A:=\F[x]/_{\DS
     \ideal{x^n-1}}=\Big\{\sum_{i=0}^{n-1}f_ix^i\;
       \mod(x^n-1)\,\Big|\,f_0,\ldots,f_{n-1}\in\F\Big\},
\end{equation}
identified with $\F^n$ in the canonical way via
\begin{equation}\label{e-p}
  \p: \F^n\longrightarrow A,\quad
  (v_0,\ldots,v_{n-1})\longmapsto\sum_{i=0}^{n-1}v_ix^i.
\end{equation}
In order to extend this situation to the convolutional setting,
we have to replace the vector space~$\F^n$ by the free module 
$\F[z]^n:=\{\sum_{\nu=0}^Nz^{\nu}v_{\nu}\mid N\in\N_0,\,v_{\nu}\in\F^n\}$ and,
consequently, the ring~$A$ by the polynomial ring
\[
   A[z]:=\Big\{\sum_{\nu=0}^Nz^{\nu}a_{\nu}\,\Big|\, N\in\N_0,\,a_{\nu}\in A\Big\}
\]
over~$A$.
Then we can extend the mapping~$\p$ above coefficientwise to polynomials, thus
$\p\big(\sum_{\nu=0}^Nz^{\nu}v_{\nu}\big)=\sum_{\nu=0}^Nz^{\nu}\p(v_{\nu})$
where, of course, $v_{\nu}\in\F^n$ and thus $\p(v_{\nu})\in A$ for all~$\nu$.
At this point it is quite natural to declare a \CC\ $\cC\subseteq\F[z]^n$ cyclic 
if it is invariant under the cyclic shift, i.~e., if\eqnref{e-cs} holds true for all
$(v_0,\ldots,v_{n-1})\in\F[z]^n$. 
Since, just like for block codes, the cyclic shift in $\F[z]^n$ corresponds to
multiplication by~$x$ in $A[z]$,
this amounts to the same as saying that~$\cC$ is called cyclic if 
$\p(\cC)$ is an ideal in $A[z]$.
However, it has been shown in~\cite[Thm.~3.12]{Pi76} and \cite[Thm.~6]{Ro79} that
each convolutional code that is cyclic in this sense has overall constraint length 
zero, thus is a block code. 
An elementary proof can be found at~\cite[Prop.~2.7]{GS04}.
Due to this result Piret~\cite{Pi76} introduced a different notion of
cyclicity for \CC{}s which then was further generalized by Roos~\cite{Ro79}.
This concept is based on some automorphism of the $\F$-algebra~$A$. 
Thus, let $\AutF(A)$ be the group of all $\F$-automorphisms on~$A$.
It is clear that each automorphism $\sigma\in\AutF(A)$ is uniquely
determined by the single value $\sigma(x)\in A$, but not every choice for 
$\sigma(x)$ determines an automorphism on~$A$.

The main idea of Piret was to impose a new ring structure on $A[z]$ and to declare a
code cyclic if it is a left ideal with respect to that ring structure.
The new structure is in general non-commutative and based on an (arbitrarily chosen)
$\F$-automorphism on~$A$.
In detail, this looks as follows.

\newpage
\begin{defi}\label{D-CCC}
Let $\sigma\in\AutF(A)$.
\begin{arabiclist}
\item On the set $A[z]$ we define addition as usual and multiplication via
      \begin{equation}\label{e-az}
         az=z\sigma(a) \text{ for all }a\in A
      \end{equation}
      along with associativity and distributivity, where  
      multiplication inside~$A$ is defined as usual. 
      This turns $A[z]$ into a (non-commutative) ring which is denoted by $\Azs$.
\item Consider the mapping 
      \[
          \p:\F[z]^n\longrightarrow\Azs,\quad 
	  \sum_{\nu=0}^Nz^{\nu}v_{\nu}\longmapsto\sum_{\nu=0}^Nz^{\nu}\p(v_{\nu})
      \]
      where $\p:\F^n\rightarrow A$ is as in\eqnref{e-p}.
      A submodule $\cC\subseteq\F[z]^n$ is said to be
      $\sigma$-{\em cyclic\/} if $\p(\cC)$ is a left ideal in $\Azs$.
\end{arabiclist}
\end{defi}

A few comments are in order. 
First notice that, unless~$\sigma$ is the identity, the indeterminate~$z$ does not 
commute with its coefficients.
Due to this very specific non-commutativity the ring $\Azs$ is also called a
{\em skew-polynomial ring}.
Since $\sigma|_{\F}=\text{id}_{\F}$, the classical polynomial ring 
$\F[z]$ is a commutative subring of $\Azs$, too.
As a consequence, $\Azs$ is a left and right $\F[z]$-module and it can easily be 
seen that the mapping~$\p$ in~(2) above is an isomorphism of left 
$\F[z]$-modules.
In the special case where $\sigma=\text{id}_{A}$ the ring $\Azs$ is
the classical commutative polynomial ring and due to the results mentioned earlier
this does not result in any convolutional codes other than block codes.
In many cases where~$\sigma$ is not the identity there do indeed exist cyclic 
convolutional codes with positive overall constraint length.
Characterizations along with several examples (actually all optimal with respect 
to their free distances) have been presented in \cite{GS04,GS03,GL03}.
Another class of such codes is given by 
some of the codes of Theorem~\ref{T-MDS}. Indeed, we have the following.

\begin{prop}\label{P-cyclic}
Let $n$ be a positive integer coprime with $|\F|$ and let 
$\alpha\in\F$ be such that $\ord(\alpha)=n$.
Furthermore, let $\delta\in\N_0$ and~$G$ be as in\eqnref{e-Gmatrix}.
Then the $\F$-algebra homomorphism $\sigma: A\longrightarrow A$ defined by $\sigma(x)=\alpha x$ is an
$\F$-automorphism on~$A$ and the submodule $\cC=\im G$ is $\sigma$-cyclic.
\\
In particular, if $\delta<n$, then $\cC$ is a cyclic MDS convolutional code.
\end{prop}
Remember that only for specific values of~$\delta$ these submodules are actually 
convolutional codes, i.~e., the matrix~$G$ is right-invertible, see also 
Remark~\ref{R-largedelta}.
Recall also from the last section that in the case $\ord(\alpha)=n>\delta$ the codes 
can be described by a certain Vandermonde parity check matrix. 
Therefore, in this case the codes have quite a rich structure.

{\sc Proof:} 
First of all, 
since $(\alpha x)^i,i=0,\ldots,n-1$ are linearly independent over~$\F$ and 
$(\alpha x)^n=1$, the mapping~$\sigma$ as defined above is indeed an automorphism 
on~$A$.
Consider now the submodule $\cC=\im G$ where 
$G=\sum_{\nu=0}^{\delta}z^\nu
   \begin{pmatrix}1&\alpha^\nu&\alpha^{2\nu}&\ldots&\alpha^{(n-1)\nu}\end{pmatrix}$.
We have to prove that $\p(\cC)$ is a left ideal in $\Azs$. 
Since $\p(\cC)$ is a left $\F[z]$-module, it suffices to show that $\p(\cC)$ is closed with
respect to left multiplication by~$x$.
Thus, consider the image of~$G$ under the mapping~$\p$, i.~e., define
\begin{equation}\label{e-g}
  g:=\p(G)=
  \sum_{\nu=0}^\delta z^{\nu}\sum_{i=0}^{n-1}\alpha^{\nu i}x^i\in\Azs.
\end{equation}
Now it suffices to show that $\p^{-1}(xg)\in \im G$, or, stated differently, that~$xg$ is 
a left $\F[z]$-multiple of~$g$. But this can easily be seen since
$$
  xg=\sum_{\nu=0}^{\delta}z^{\nu}\sigma^{\nu}(x)\sum_{i=0}^{n-1}\alpha^{\nu i}x^i\\
    =\sum_{\nu=0}^{\delta}z^{\nu}\sum_{i=0}^{n-1}\alpha^{\nu(i+1)}x^{i+1}
    =\sum_{\nu=0}^\delta z^{\nu}\sum_{i=0}^{n-1}\alpha^{\nu i}x^i=g.
  \eqno\Box
$$

This result proves in particular that if $\ord(\alpha)=n$, then the codes in 
Theorem~\ref{T-MDS} are cyclic.
In the sequel we want to show even more. 
Indeed, we will prove that the condition $\ord(\alpha)=n$ is even necessary and 
sufficient for the codes of Theorem~\ref{T-MDS} to be cyclic with respect to some 
$\F$-automorphism.

To this end we need some more details about the coefficient ring $A$.
Due to the coprimeness of~$n$ and~$|\F|$, this ring is a direct product of fields. 
Indeed, let
\begin{equation}\label{e-xn-1}
        x^n-1=\pi_1\cdot\ldots\cdot \pi_r,
\end{equation}
where $\pi_1,\ldots,\pi_r\in\F[x]$ are irreducible, monic, and pairwise different.
Then the Chinese Remainder Theorem tells us that
\begin{equation}\label{e-CRT}
  \psi: A\longrightarrow K_1\times\ldots\times K_r,\quad
         a\longmapsto \big(a~\mod\,\pi_1,\ldots,a~\mod\,\pi_r\big),
\end{equation}
where $K_k=\F[x]/_{\DS\ideal{\pi_k}}$, is an isomorphism if $\times_{i=1}^rK_i$ 
is endowed with componentwise addition and multiplication.
Notice that $K_k\cong K_l$ if and only if $\deg_x\pi_k=\deg_x\pi_l$.
The elements 
\begin{equation}\label{e-epsk}
     \ve{k}:=\psi^{-1}\big(0,\ldots,0,1,0,\ldots,0\big) 
     \text{ for } k=1,\ldots,r
\end{equation}
(where~$0$ and~$1$ have to be understood as the elements 
$0\,\mod\,\pi_l$ and $1\,\mod\,\pi_l$ in~$K_l$)
are particularly important since they form the uniquely determined set of primitive 
idempotents in~$A$.
The idempotents are pairwise orthogonal, thus $\ve{k}\ve{l}=0$ for 
$k\not=l$.
Observe that for any $a\in A$ the products $\ve{l}a$ single out the various components 
of~$a$. Precisely, 
$\psi(\ve{l}a)=(0,\ldots,0,a\,\mod\,\pi_l,0,\ldots,0)$ for any $l=1,\ldots,r$.
Therefore, $\ve{1}a+\ldots+\ve{r}a$ is a decomposition of $a\in A$ just like 
the one in\eqnref{e-CRT} and in the sequel we will use this representation rather than
that from\eqnref{e-CRT}.

It is straightforward to see that a given automorphism $\sigma\in\AutF(A)$ induces a 
permutation on the set of primitive idempotents. More precisely, 
\begin{equation}\label{e-sigmaeps}
   \sigma(\ve{k})=\ve{l} \text{ for some $l$ such that $\deg_x\pi_k=\deg_x\pi_l$.}
\end{equation}

The following example will be important for our purposes.

\begin{exa}\label{E-ord(alpha)n}
Let $\alpha\in\F$ be such that $\ord(\alpha)=n$. Then $x^n-1$ decomposes into linear 
factors, precisely,
\begin{equation}\label{e-xnn-1}
  x^n-1=(x-1)(x-\alpha)\cdot\ldots\cdot(x-\alpha^{n-1}).
\end{equation}
Along with the~$n$ irreducible factors, we also have~$n$ idempotents.  
We will denote them by $\ve{0},\ldots,\ve{n-1}$.
Due to\eqnref{e-epsk} they have to satisfy $\ve{k}(\alpha^i)=\delta_{k,i}$ for all
$k,\,i=0,\ldots,n-1$.
Thus, the idempotents are of the form
\[
   \ve{k}=\gamma_k\prod_{i=0\atop i\not=k}^{n-1}(x-\alpha^{i})\ 
   \text{ for some }\gamma_k\in\F^*,\quad k=0,\ldots,n-1.
\]
In particular, $\ve{0}=\gamma_0\frac{x^n-1}{x-1}=\gamma_0\sum_{i=0}^{n-1}x^i$ and 
$\ve{0}(1)=1$ shows that $\gamma_0=\frac{1}{n}$, which indeed exists in~$\F^*$ since~$n$
and~$|\F|$ are coprime.
Consider now the automorphism $\sigma\in\AutF(A)$ defined by $\sigma(x)=\alpha x$, see 
Proposition~\ref{P-cyclic}.
Then, using $\ord(\alpha)=n$, we obtain
\[
  \sigma(\ve{k})=\ve{k}(\alpha x)
   =\gamma_k\prod_{i=0\atop i\not=k}^{n-1}(\alpha x-\alpha^{i})
   =\gamma_k\alpha^{n-1}\prod_{i=0\atop i\not=k}^{n-1}(x-\alpha^{i-1})
   =\gamma_k\alpha^{n-1}\prod_{i=0\atop i\not=k-1}^{n-1}(x-\alpha^{i}).
\]
Since~$\sigma(\ve{k})$ is one of the idempotents again, see\eqnref{e-sigmaeps},
it follows $\sigma(\ve{k})=\ve{k-1}$ for $k=0,\ldots,n-1$, where we take exponents 
modulo~$n$.
In particular, $\sigma^{\nu}(\ve{0})=\ve{n-\nu}$. 
Hence~$\sigma$ induces the permutation with cycle notation
\begin{equation}\label{e-cycle}
   \Big(\ve{n-1},\ve{n-2},\ldots,\ve{1},\ve{0}\Big).
\end{equation}
Using $\sigma^{\nu}(x)=\alpha^{\nu}x$ we obtain from the above
\[
  \ve{n-\nu}=\sigma^{\nu}(\ve{0})=\ve{0}(\alpha^{\nu}x)
  =\frac{1}{n}\sum_{i=0}^{n-1}(\alpha^{\nu}x)^i \text{ for }\nu\geq 0.
\]
This shows that the polynomial~$g$ in\eqnref{e-g} satisfies
\begin{equation}\label{e-gg}
  g=n\sum_{\nu=0}^{\delta} z^{\nu}\sigma^{\nu}(\ve{0})
   =n\sum_{\nu=0}^{\delta}z^{\nu}\ve{n-\nu}=n\ve{0}\sum_{\nu=0}^{\delta}z^{\nu}.
\end{equation}
We will make use of this representation later on.
\end{exa}

Now we can prove that the codes in Theorem~\ref{T-MDS} are cyclic if and only if 
$\ord(\alpha)=n$.
Just like in Proposition~\ref{P-cyclic} we will consider arbitrary overall
constraint length~$\delta$.
However, the case $\delta=0$ needs to be excluded since it gives us, for any order of~$\alpha$, 
a cyclic (block) code. 
We will make heavy use of the results derived in~\cite{GS04}.

\begin{theo}\label{T-cyclic}
Let $n\in\N$ be such that~$n$ and~$|\F|$ are coprime and let $\alpha\in\F$ be such that
$\ord(\alpha)\geq n$. 
Moreover, let $\delta\in\N$ and put~$G$ as in\eqnref{e-Gmatrix}. 
Define $\cC:=\im G$. 
Then~$\cC$ is $\sigma$-cyclic for some $\sigma\in\AutF(A)$ if and only if $\ord(\alpha)=n$.
In this case~$\cC$ is $\sigma$-cyclic for the automorphism $\sigma\in\AutF(A)$ defined via 
$\sigma(x)=\alpha x$.
\end{theo}

\begin{proof}
The if-part as well as the additional statement have been proven in Proposition~\ref{P-cyclic}. 
\\
``Only-if-part'': 
Let $\cC=\im G$ be $\sigma$-cyclic for some 
$\sigma\in\AutF(A)$.
Let $x^n-1=\prod_{i=1}^r\pi_i$ be the prime factorization of~$x^n-1$ such that 
$\pi_1=x-1$. 
Denoting the idempotents by $\ve{1},\ldots,\ve{r}$ we have in particular 
$\ve{1}=\frac{1}{n}\frac{x^n-1}{x-1}=\frac{1}{n}\sum_{i=0}^{n-1}x^i$.
By assumption $\ord(\alpha)\geq n$ and we have to show that $\ord(\alpha)=n$, hence that 
$r=n$ and, up to ordering, $\pi_i=x-\alpha^{i-1}$ for $i=1,\ldots,n$.
By assumption, $\p(\cC)$ is the left ideal generated by~$g$ as given in\eqnref{e-g}.
Since $\rank\cC=1$, the generator matrix~$G$ is unique up to a nonzero constant in~$\F$.
Thus, the generator of the left ideal is also unique up to a constant factor and, 
along with \cite[Cor.~4.13 and Thm.~4.15(b)]{GS04}, this shows that the polynomial~$g$ is 
reduced in the sense of \cite[Def.~4.9(b)]{GS04}.
Moreover, since the code is one-dimensional we obtain from \cite[Thm.~7.13]{GS04} that 
$g=\ve{k}g$ for some $k=1,\ldots,r$ such that $\deg\pi_k=1$.
In particular we have $g_0=\ve{k}g_0$ for the constant coefficient~$g_0$ of~$g$.
Since\eqnref{e-g} yields $g_0=n\ve{1}$ and $\ve{k}\ve{1}=0$ for $k>1$, we conclude $k=1$, 
thus $g=\ve{1}g$.
Therefore,
\[
   g_0+zg_1+z^2g_2+\ldots+z^{\delta}g_{\delta}
   =\ve{1}g_0+z\sigma(\ve{1})g_1+z^2\sigma^2(\ve{1})g_2+\ldots+
          z^{\delta}\sigma^{\delta}(\ve{1})g_{\delta}
\]
where $g_{\nu}$ is the coefficient of~$z^{\nu}$ in~$g$. 
Hence, $g_{\nu}=\sigma^{\nu}(\ve{1})g_{\nu}$ for all $\nu=0,\ldots,\delta$.
Moreover, since $\delta>0$ we have $\sigma(\ve{1})\not=\ve{1}$ for otherwise the code 
would have overall constraint length zero, see \cite[Lemma~3.4]{GL03}.
Consider now the coefficient $g_1=\sum_{i=0}^{n-1}(\alpha x)^i$.
The equation $g_1=\sigma(\ve{1})g_1$ along with $\sigma(\ve{1})\not=\ve{1}$ and the 
orthogonality of the idempotents implies 
$\ve{1}g_1=0$.
Substituting $x=1$, we obtain $\sum_{i=0}^{n-1}\alpha^i=0$.
But then $\sum_{i=0}^{n-1}\alpha^i(\alpha-1)=\alpha^n-1=0$ which along with the assumption
$\ord(\alpha)\geq n$ implies $\ord(\alpha)=n$.
\end{proof}

We want to close the paper with yet another representation of the cyclic codes considered 
so far.
In~\cite[Prop.~7.10]{GS04} it has been shown that a polynomial~$g\in\Azs$ with the property
$g=\ve{k}g$ for some $k=1,\ldots,r$ generates an ideal that is a convolutional code, i.~e., 
a direct summand in the left $\F[z]$-module $\Azs$, if and only if $g=\ve{k}u$ for some
unit $u\in\Azs$.
More details about this can be found in~\cite{GL03}.
From Example~\ref{E-ord(alpha)n} we can easily derive how such a unit looks like in the 
case of the equivalent conditions of Theorem~\ref{T-cyclic} if~$\delta$ is not too big.

\begin{prop}\label{P-unit}
Let $\ord(\alpha)=n$ and let $\sigma\in\AutF(A)$ be defined via $\sigma(x)=\alpha x$. 
Furthermore, let $1\leq\delta\leq n$.
Let $x^n-1$ be factored as in\eqnref{e-xnn-1} and denote the 
corresponding idempotents by $\ve{0},\ldots,\ve{n-1}$.
Then the polynomial~$g$ from\eqnref{e-g} satisfies $g=\ve{0}u$ where 
\[
  u=n(1+z\ve{n-1})(1+z\ve{n-2})\cdot\ldots\cdot(1+z\ve{n-\delta})
\]
and~$u$ is a unit in $\Azs$. 
\end{prop}

\begin{proof}
First of all we have for each $j=1,\ldots,n$ 
\[
   (1+z\ve{n-j})(1-z\ve{n-j})=(1-z\ve{n-j})(1+z\ve{n-j})=1,
\]
since $\sigma(\ve{n-j})=\ve{n-j-1}$ due to\eqnref{e-cycle}, and since the idempotents
are pairwise orthogonal.
Thus~$u$ is indeed a unit in $\Azs$.
Moreover, using the identity $\sigma(\ve{k})=\ve{k-1}$, see\eqnref{e-cycle}, and 
again the orthogonality of the idempotents one can show by induction on~$\delta$ that
\[
  u=n\Big(1+z\sum_{k=n-\delta}^{n-1}\ve{k}+z^2\sum_{k=n-\delta}^{n-2}\ve{k}+\ldots+
    z^{\delta}\sum_{k=n-\delta}^{n-\delta}\ve{k}\Big)
    \text{ for }\delta=1,\ldots,n,
\]
where a sum is zero if the lower index is strictly bigger than the upper one.
From this and\eqnref{e-gg} one can easily see that $\ve{0}u=g$.
\end{proof}

Since the element $n\in\F$ is a unit in $\Azs$ we can summarize the results of 
Example~\ref{E-ord(alpha)n} and the previous proposition as follows, see in 
particular\eqnref{e-gg} and also\eqnref{e-g}.

\begin{theo}\label{T-cyclicsumm}
Let $\ord(\alpha)=n$ and $\sigma\in\AutF(A)$ be such that $\sigma(x)=\alpha x$. 
Let~$\cC$ be as in Theorem~\ref{T-MDS}. Then
\begin{alphalist}
\item $\p(\cC)$ is the left ideal in $\Azs$ generated by the element 
      \[
        \sum_{\nu=0}^\delta z^{\nu}\sum_{i=0}^{n-1}\alpha^{\nu i}x^i=
	\prod_{i=1}^{n-1}(x-\alpha^i)\sum_{\nu=0}^\delta z^{\nu}.
      \]
\item $\p(\cC)$ is the left ideal generated by the element 
      \[
         \ve{0}\sum_{\nu=0}^\delta z^{\nu}=
	 \ve{0}(1+z\ve{n-1})(1+z\ve{n-2})\cdot\ldots\cdot(1+z\ve{n-\delta}).
      \]
\end{alphalist}
The representation on the right hand side of~(a) justifies to call these codes one-dimensional
Reed-Solomon convolutional codes.
\end{theo}
In the paper~\cite{GL03} representations of cyclic codes via units like on the right hand side of~(b)
above have been studied in detail.
Therein, it has been investigated as to which algebraic parameters (field size, dimension, 
overall constraint length, and Forney indices) can be realized by cyclic convolutional 
codes. 
In particular, a construction of certain compact cyclic convolutional codes (i.~e., all 
Forney indices are the same) has been derived.
However, no distance results have been obtained in that context.
As has been shown in~\cite{GS04a} the presentation as on the right hand side of~(a) seems to be more 
suitable for a generalization to codes of higher dimension with good distance.

\section*{Open Problems}
We have presented a class of one-dimensional convolutional codes with maximum 
possible distance. In the specific case where $\ord(\alpha)=n$ 
these codes are cyclic and 
have a Vandermonde parity check matrix. 
Without using explicitly Vandermonde matrices, but highly the theory of
cyclic convolutional codes, first attempts are currently under investigation 
of how to generalize the construction of cyclic convolutional codes with large distance 
to higher dimensions, see~\cite{GS04a}.
In general, we consider it most important to understand whether Vandermonde structure of a 
cyclic convolutional code can be exploited for distance computations and algebraic 
decoding algorithms.
We think that the one-dimensional cyclic MDS codes with their rich structure as 
presented in this paper might be a good starting point in this regard.

\bibliographystyle{abbrv}
\bibliography{literatureAK,literatureLZ}
\end{document}